\documentclass[english,12pt]{article}
          \usepackage{babel}
          \usepackage[T1]{fontenc}
          \usepackage[latin2]{inputenc}
          \usepackage{pslatex}
\begin{document}

\title{\bf Curious Variables Experiment (CURVE). \\
Superhump Period Change Pattern in KS UMa and Other Dwarf Novae}
\author{A. ~O~l~e~c~h$^1$, ~A. ~S~c~h~w~a~r~z~e~n~b~e~r~g~-~C~z~e~r~n~y~$^{1,3}$,\\
~P. ~K~\c{e}~d~z~i~e~r~s~k~i$^2$, ~K. ~Z~{\l}~o~c~z~e~w~s~k~i$^2$, ~K. ~M~u~l~a~r~c~z~y~k~$^2$,
\\~and~ \\
M. ~W~i~\'s~n~i~e~w~s~k~i$^2$}
\date{$^1$ Nicolaus Copernicus Astronomical Center, 
Polish Academy of Sciences,
ul.~Bartycka~18, 00-716~Warszawa, Poland,\\ 
{\tt e-mail: (olech,alex,jis)@camk.edu.pl}\\
~\\
$^2$ Warsaw University Observatory, Al. Ujazdowskie 4, 00-476 Warszawa, 
Poland, {\tt e-mail: (kmularcz,kzlocz,pkedzier,mwisniew)@astrouw.edu.pl}\\
~\\
$^3$ Adam Mickiewicz University Observatory, ul. S{\l}oneczna 36,
60-286 Poznan, Poland
}

\maketitle

\begin{abstract}

We report extensive photometry of the dwarf nova KS UMa throughout its
2003 superoutburst till quiescence. During the superoutburst the star
displayed clear superhumps with a mean period of $P_{sh} = 0.070092(23)$
days. In the middle stage of superoutburst the period was increasing
with a rate of $\dot P/P = (21\pm12)\times 10^{-5}$ and later was
decreasing with a rate of $\dot P/P = -(21\pm8)\times 10^{-5}$.

At the end of superoutburst and during first dozen days of quiescence
the star was showing late superhumps with a mean period of $P_{late} =
0.06926(2)$ days. This phenomenon was observed even 30 days after
beginning of the superoutburst.

In quiescence the star shows quasi-periodic modulations with amplitude
reaching 0.5 mag. The most common structure observed during this stage
was sinusoidal wave characterized by a period of about 0.1 days.

Comparing KS UMa to other SU UMa stars we conclude that this group of
dwarf novae shows decreasing superhump periods at the beginning and the
end of superoutburst but increasing period in the middle phase.

\end{abstract}

\section{Introduction}

Balayan (1997) identified 10 cataclysmic variables among the stars
of the Second Buryakan Sky Survey (Makarian and Stepanian 1983). Seven
of these variables were already known, but three were new. One of these
new objects was KS UMa (SBS 1017+533).

\vspace{7.5cm}

\includegraphics{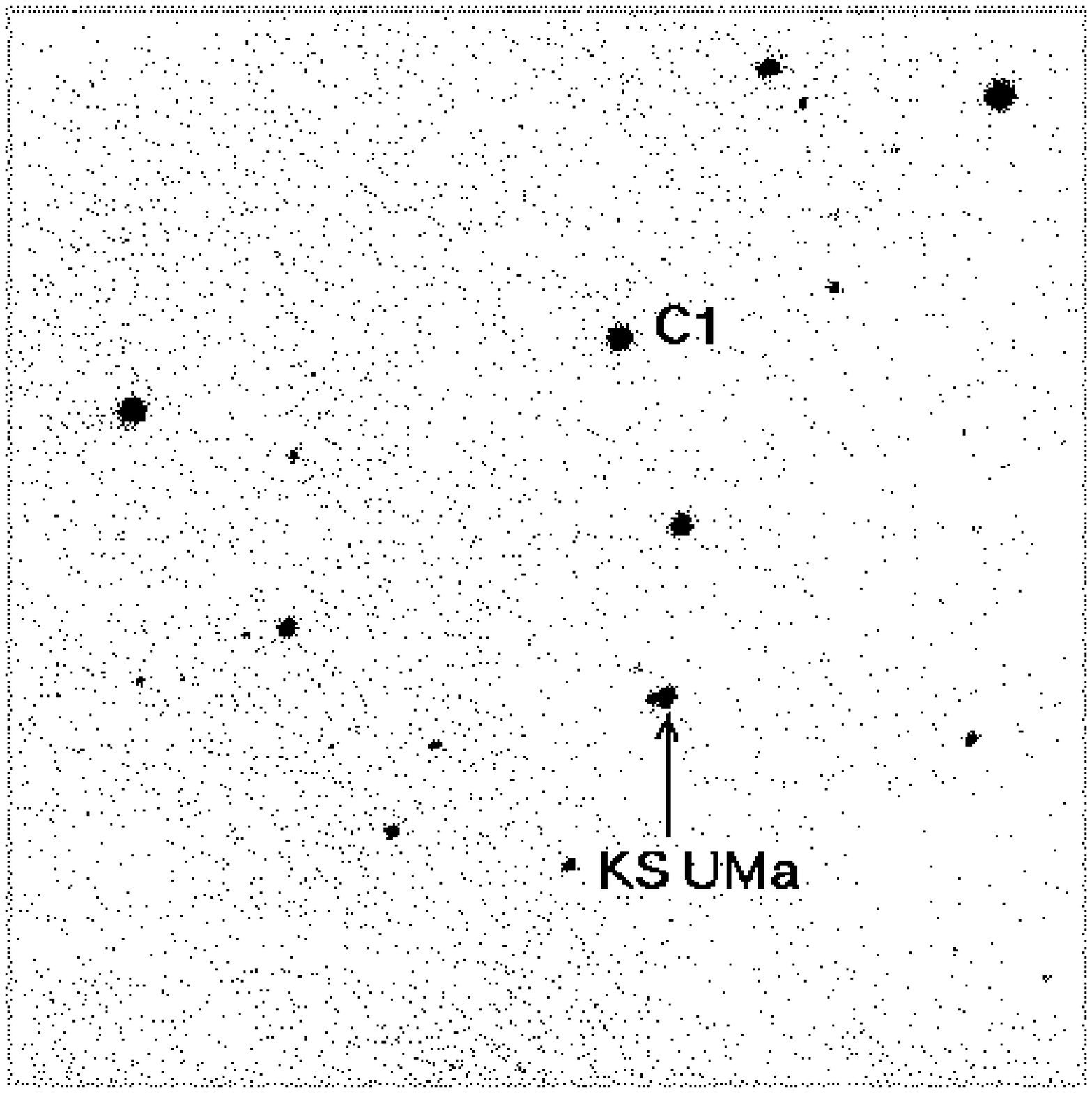}

\begin{figure}[h]
\caption{\sf Finding chart for KS UMa covering a region of $6.5 \times
6.5$ arcminutes. The position of the comparison star is shown. North is
up, east is left.}
\end{figure}

\vspace{8.5cm}

\includegraphics{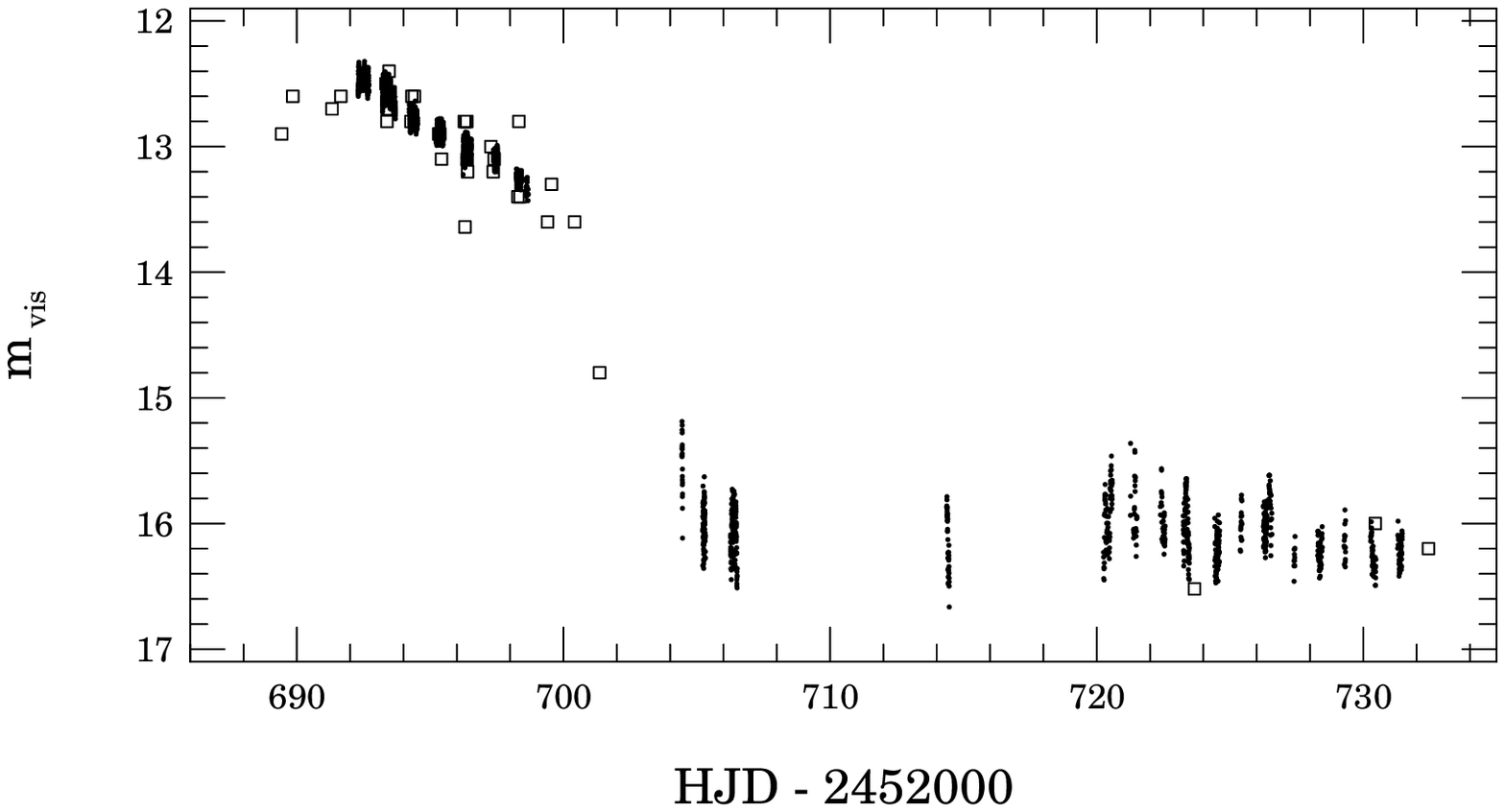}

\begin{figure}[h]
\caption{\sf The general photometric behavior of KS UMa during
its 2003 superoutburst. Visual estimates collected in the VSNET archive
are shown as open squares and CCD observations described in this
paper as dots.}
\end{figure}
\bigskip

In 1998 the star was observed in the bright state. Detection of the
superhumps with period of 0.0697 day during this event by T. Vanmunster
\footnote {VSNET alert no. 1448} proved that KS UMa belongs to the
group of SU UMa type dwarf novae.

Historical light curve of the star based on the Harvard College
Observatory photographic plates was obtained by Hazen and Garnavich
(1999).

KS UMa most probably coincides with the ROSAT X-ray source J1020.4+5304
(Snowden et al. 1995) located only 10 arc sec away from the variable.

On 2003 February 18/19 KS UMa went into the superoutburst again as was
announced by Eddy Muyllaert and Gary Poyner. \footnote
{VSNET-superoutburst 1919 alert} Because of the excellent visibility,
and because of the lack of the entire coverage of a superoutburst in the
past, we performed extensive CCD photometry of the star both in the
superoutburst and quiescence.

\section{Observations}

Observations of KS UMa reported in the present paper were obtained during
23 nights from 2003 February 21 to April 1 at the
Ostrowik station of the Warsaw University Observatory. They were
collected using the 60-cm Cassegrain telescope equipped with a
Tektronics TK512CB back illuminated CCD camera. The scale of the camera
was 0.76"/pixel providing a $6.5'\times 6.5'$ field of view. The full
description of the telescope and camera was given by Udalski and Pych
(1992).

We monitored the star in ``white light''. This was due to the lack of
an autoguiding system, not yet implemented after recent telescope
renovation. Thus we did not use any filter to shorten the exposures in
order to minimize guiding errors. 

The exposure times were from 30 to 90 seconds during the bright state
and from 150 to 300 seconds in the minimum light.

A full journal of our CCD observations of KS UMa is given in Table
1. In total, we monitored the star during 110.18 hours and obtained 3150
exposures.

\begin{table}[h]
\caption{\sc Journal of the CCD observations of KS UMa}
\vspace{0.1cm}
\begin{center}
\begin{tabular}{|l|c|c|r|r|}
\hline
\hline
Date of& Time of start & Time of end & Length of & Number of \\
2003   & 2452000. + & 2452000. + & run [hr]~ & frames \\
\hline
Feb. 21/22 &692.30016 &692.69672 & 6.313 & 284\\
Feb. 22/23 &693.21151 &693.69467 &11.596 & 624\\
Feb. 23/24 &694.22142 &694.52265 & 7.230 & 365\\
Feb. 24/25 &695.21588 &695.49997 & 6.818 & 285\\
Feb. 25/26 &696.23725 &696.54021 & 7.271 & 425\\
Feb. 26/27 &697.37482 &697.52844 & 3.687 & 219\\
Feb. 27/28 &698.23340 &698.67746 & 6.931 & 139\\
Mar. 05/06 &704.44387 &704.46670 & 0.548 &  20\\
Mar. 06/07 &705.22579 &705.31920 & 2.242 &  75\\
Mar. 07/08 &706.26975 &706.51688 & 5.931 & 122\\
Mar. 15/16 &714.37474 &714.46856 & 2.252 &  45\\
Mar. 21/22 &720.25577 &720.58496 & 7.901 &  76\\
Mar. 22/23 &721.26472 &721.50796 & 3.032 &  30\\
Mar. 23/24 &722.38846 &722.56662 & 4.276 &  36\\
Mar. 24/25 &723.24628 &723.47508 & 5.491 &  78\\
Mar. 25/26 &724.42984 &724.60743 & 4.262 &  62\\
Mar. 26/27 &725.38228 &725.44741 & 1.563 &  16\\
Mar. 27/28 &726.25305 &726.57153 & 7.644 &  92\\
Mar. 28/29 &727.39841 &727.43791 & 0.948 &  12\\
Mar. 29/30 &728.28647 &728.45990 & 4.162 &  44\\
Mar. 30/31 &729.28010 &729.33608 & 1.344 &  16\\
Mar. 31/01 &730.27469 &730.46936 & 4.672 &  39\\
Apr. 01/02 &731.29169 &731.46109 & 4.066 &  46\\
\hline
Total          &   --   & -- & 110.18 & 3150 \\ 
\hline
\hline
\end{tabular}
\end{center}
\end{table}

\subsection{Data Reduction}

All the data reductions were performed using a standard procedure
based on the IRAF \footnote{IRAF is distributed by the National Optical
Astronomy Observatory, which is operated by the Association of
Universities for Research in Astronomy, Inc., under a cooperative
agreement with the National Science Foundation.} package and
the profile photometry has been derived using the DAOphotII package
(Stetson 1987).

Relative unfiltered magnitudes of KS UMa were determined as the
difference between the magnitude of the variable and the magnitude of
the comparison star GSC 3815:610 ($R.A.=10^h20^m28.25^s$, $Decl. =
+53^\circ06'36.6"$) located 2.2' to the north of the variable. This
comparison star is marked in the chart displayed in Fig. 1. 

Note that a faint companion of magnitude $V\approx18$ located few
seconds to the east of the variable does not affect our profile
photometry in any substantial way. 

The typical accuracy of our measurements varied between 0.001 and 0.012
mag in the bright state and between 0.006 and 0.055 mag in the minimum
light. The median value of the photometry errors was 0.005 and 0.014
mag, respectively.

\subsection{Light curves}

Fig. 2 presents the photometric behavior of KS UMa as observed
between February and April 2003. Relative differential magnitudes of the
variable were transformed to the visual magnitudes using visual
estimates of the brightness of the star made by astronomy amateurs and
published in VSNET archive (these observations are marked using open
squares). Photographic magnitude of our comparison star, according to
GSC catalogue is $14.59\pm0.21$ mag. With respect to that the visual
scale on Fig. 2 appears to by shifted by 0.3 mag down.

The superoutburst of KS UMa started on Feb. 18/19.\footnote{
VSNET-superoutburst 1919 alert} Subsequent CCD photometry made by Tonny
Vanmunster showed no short term modulations apart from general
brightening. \footnote{VSNET-superoutburst 1921 and 1923 alerts}
Full development of the superhumps took place on Feb. 19.\footnote{
VSNET-superoutburst 1927 alert}

\clearpage
~

\vspace{21.2cm}

\includegraphics{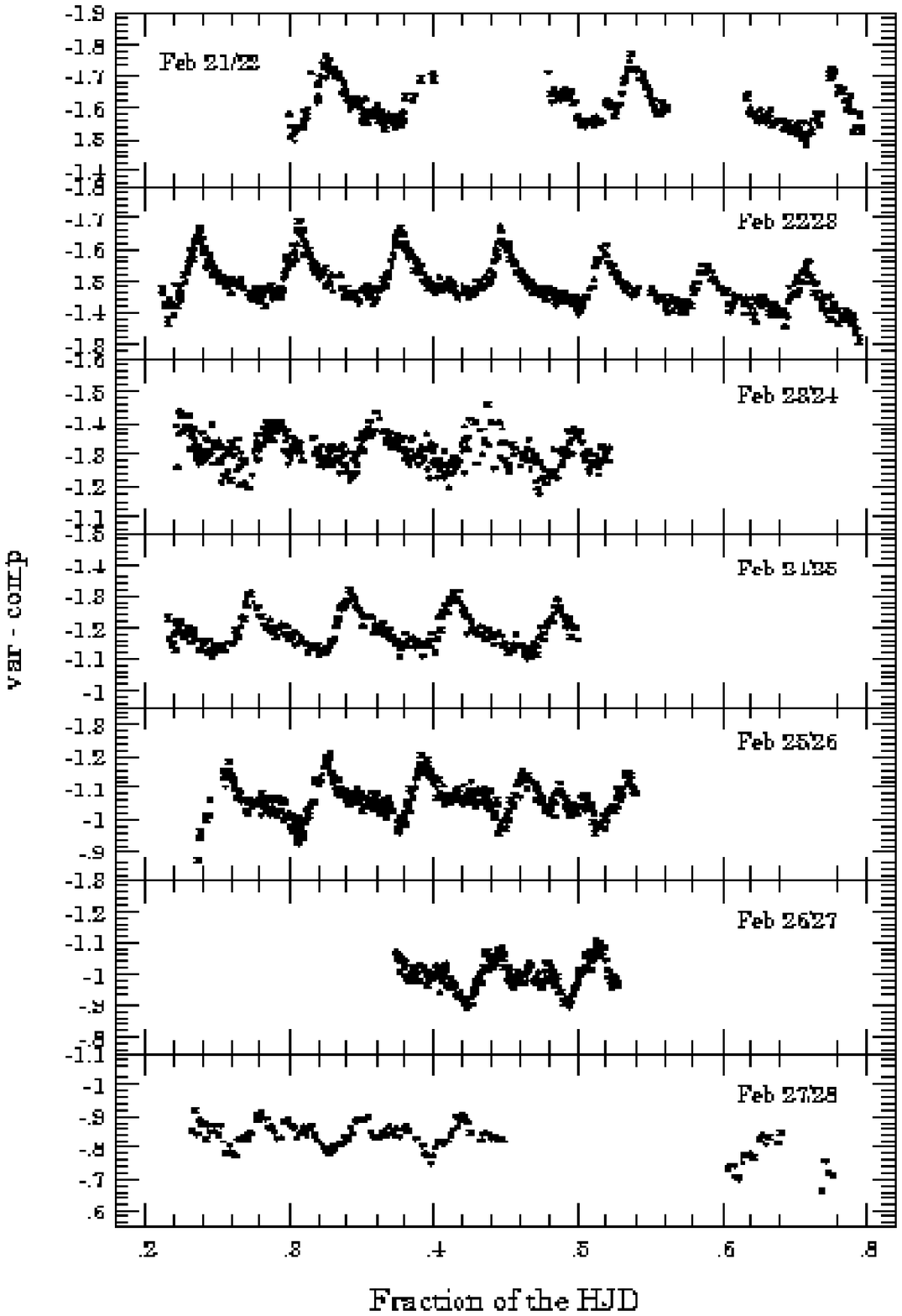}

\begin{figure}[h]
\caption{\sf The light curves of KS UMa observed during seven
consecutive nights in February 2003.}
\end{figure}
\clearpage
~

\vspace{21.2cm}

\includegraphics{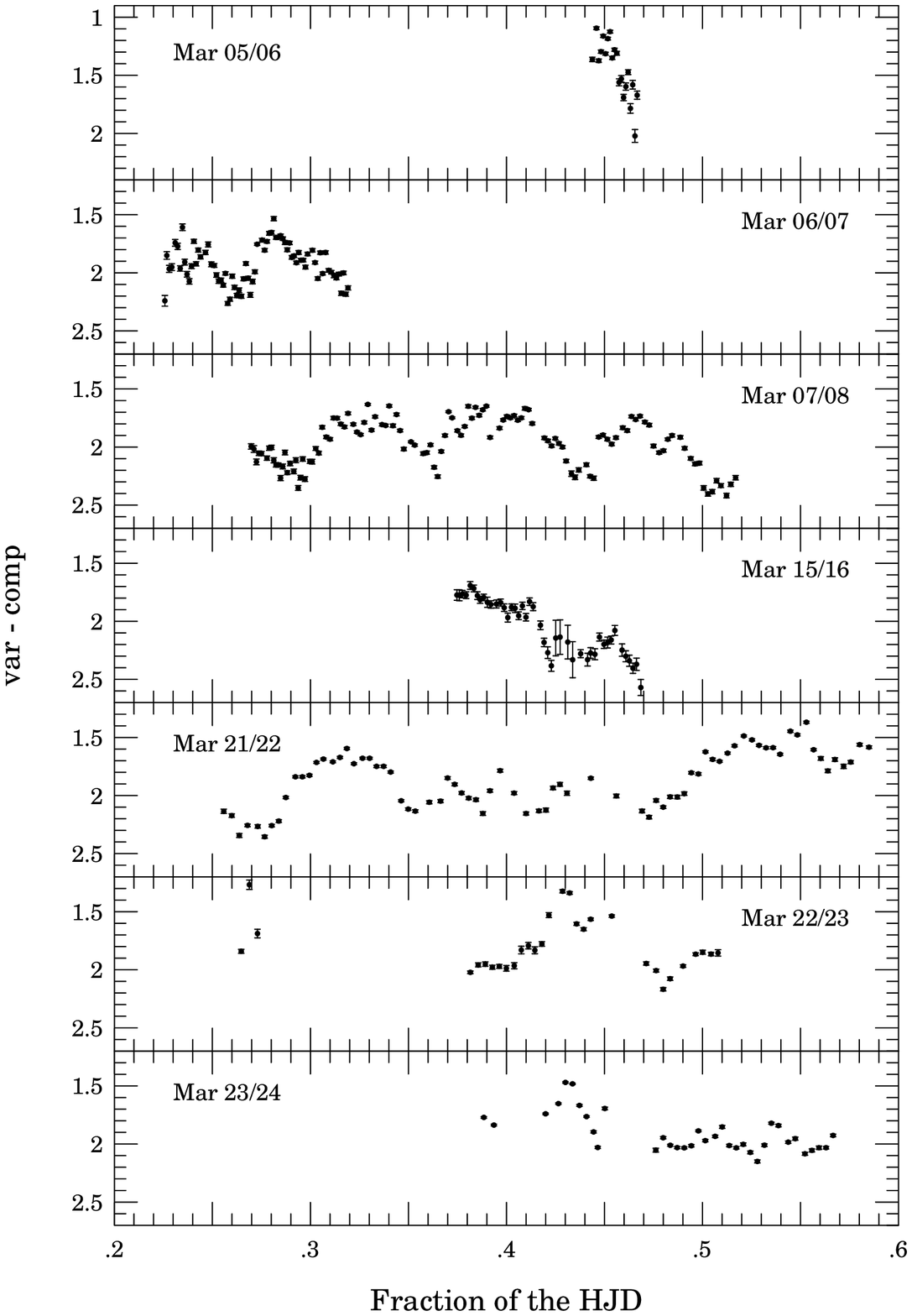}

\begin{figure}[h]
\caption{\sf The light curves of KS UMa observed between Mar. 05/06 and
Mar. 23/24.}
\end{figure}
\clearpage
~

\vspace{21.2cm}

\includegraphics{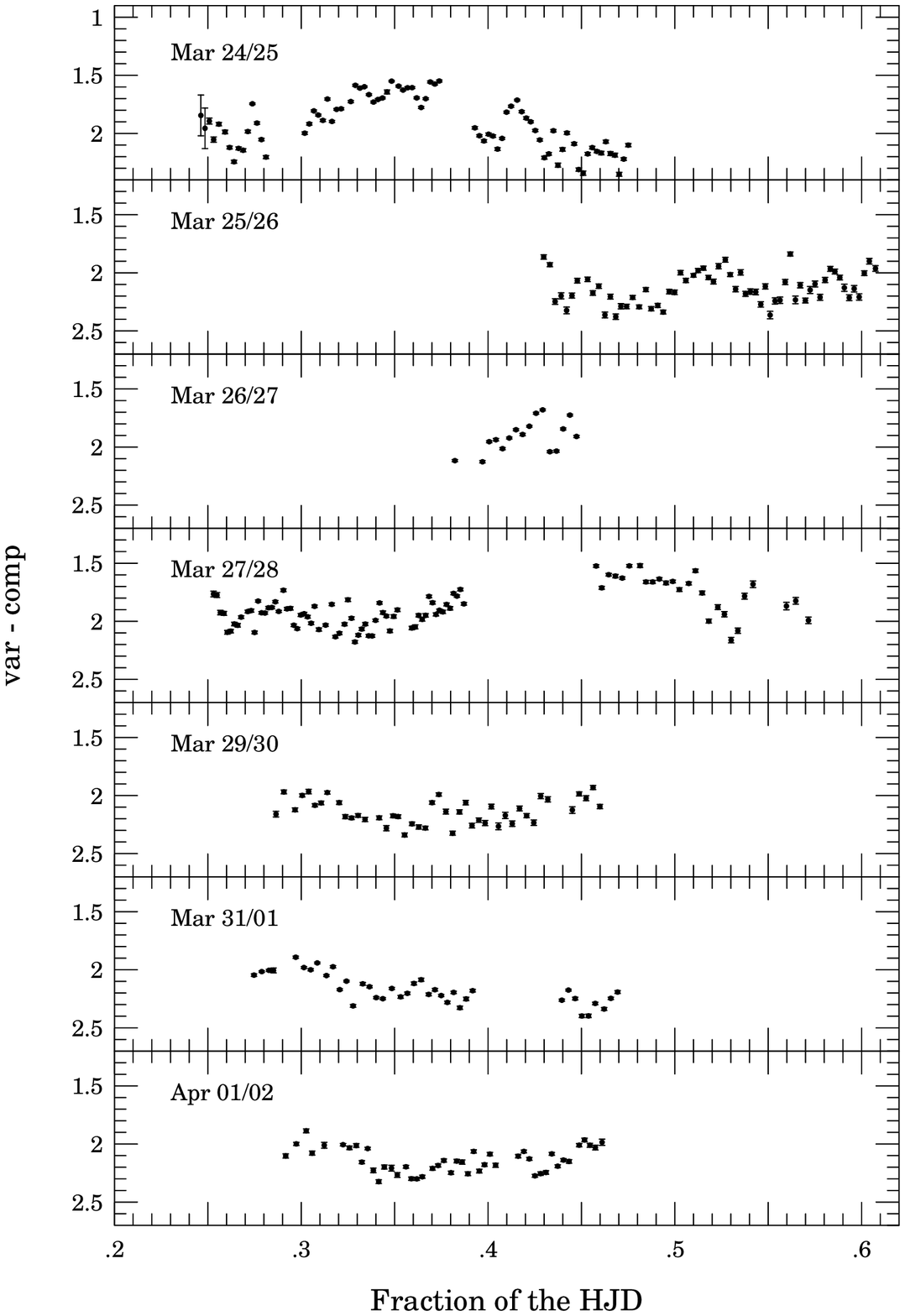}

\begin{figure}[h]
\caption{\sf The light curves of KS UMa observed between Mar. 24/25 and
Apr. 01/02.}
\end{figure}
\clearpage

Our observations started with fourth night of the superoutburst and with
the third night of the presence of the superhumps. We observed the
superoutburst during seven consecutive nights from Feb. 21/22 to Feb.
27/28. During this time the star faded by 0.85 mag giving the mean
decrease of 0.13 mag per day. Our observing run from Mar. 05/06 shows
that the star was still 0.5 mag brighter than in the quiescence. During
the night of Mar. 06/07 this difference was only 0.1 mag. Thus we
conclude that the superoutburst lasted until Mar. 06/07 i.e. 16 days.

Fig. 3 shows the light curves of KS UMa during seven consecutive nights
of February 2003. The superhumps are clearly visible in each run.

The light curves from period Mar. 05/06 to Mar. 23/24 are shown in Fig.
4. Clear periodic light variations are present even until Mar. 22/23.

Fig. 5 shows the lights curves of seven longest runs from period Mar. 
24/25 to Apr 01/02. The magnitude of the star varied during this
interval with amplitude of 0.5 mag but the changes were more chaotic and
there is no trace of strictly periodic signal as in Figs. 3 and 4.

\section{Superhumps}

\subsection{Amplitude}

In a typical SU UMa star the amplitude of the superhumps reaches its
maximum around third day of the superoutburst. Starting from this moment
the amplitude monotonically decreases. Few days later superhumps evolve
from the tooth-shape light curve to more complicated shape showing more
scatter and secondary maxima called interpulses. Around the end of the
superoutburst low amplitude superhumps switch into late humps
characterized by the same period but phase shift of $\sim0.5$ cycles.

As was described in the VSNET e-mail list \footnote {VSNET-superoutburst
1929} the peak-to-peak amplitude of the superhumps in KS UMa reached a
maximum of 0.30 mag on Feb. 20. During the first night of our run i.e.
on Feb. 21/22 the amplitude was 0.21 mag as is clearly visible in our
Fig. 6 where we show the mean superhump profiles for each night. These
profiles were obtained by phasing our observations with the superhump 
period for each night and averaging them in 0.02 - 0.05 phase bins. 

Our longest run obtained on Feb. 22/23  lasting almost 12 hours shows
gradual change in the superhump profile. During this interval the
amplitude of the superhumps decreased from 0.20 mag (first four humps)
to 0.16 mag (last 3 humps). Additionally, the main superhump maxima
became weaker, while other peaks became stronger.

During the night of Feb. 23/24 the light curve was quite noisy with 
amplitude of the modulations equal to 0.13 mag. Surprisingly, on Feb.
24/25 the amplitude increased to 0.17 mag and the shape of the
superhumps became sharp and smooth again as during the first days of the
superoutburst.

Increase of the amplitude of the superhumps has been continued until
Feb. 25/26 when it reached its local maximum with 0.18 mag. However
the shape of the light curve during this night changed markedly in
comparison with Feb. 24/25. The interpulses and other secondary humps
were also clearly visible in the light curves from Feb. 26/27 and 27/28
and during these nights amplitude was 0.16 and 0.12 mag, respectively.

\subsection{Period}

From each light curve of KS UMa in superoutburst we removed the first or
second order polynomial and then analyzed them using {\sc anova}
statistics and two harmonics Fourier series (Schwarzenberg-Czerny 1996).
The resulting periodogram is shown in Fig. 7. The most prominent peak is
found at a frequency of $f=14.263\pm0.010$ c/d, which corresponds to the
period $P_{sh}=0.07011(5)$ days ($100.96\pm0.07$ min). The peak visible
at 7.13 c/d is a ghost of main frequency arising due to use of two
harmonics. The harmonic peak at 28.53 c/d appears to be real. The inset
in Fig. 7 shows the magnification of the power spectrum around main
frequency. Apart from this main peak and its aliases the inset shows no
other significant periodicities.

\vspace{15cm}

\includegraphics{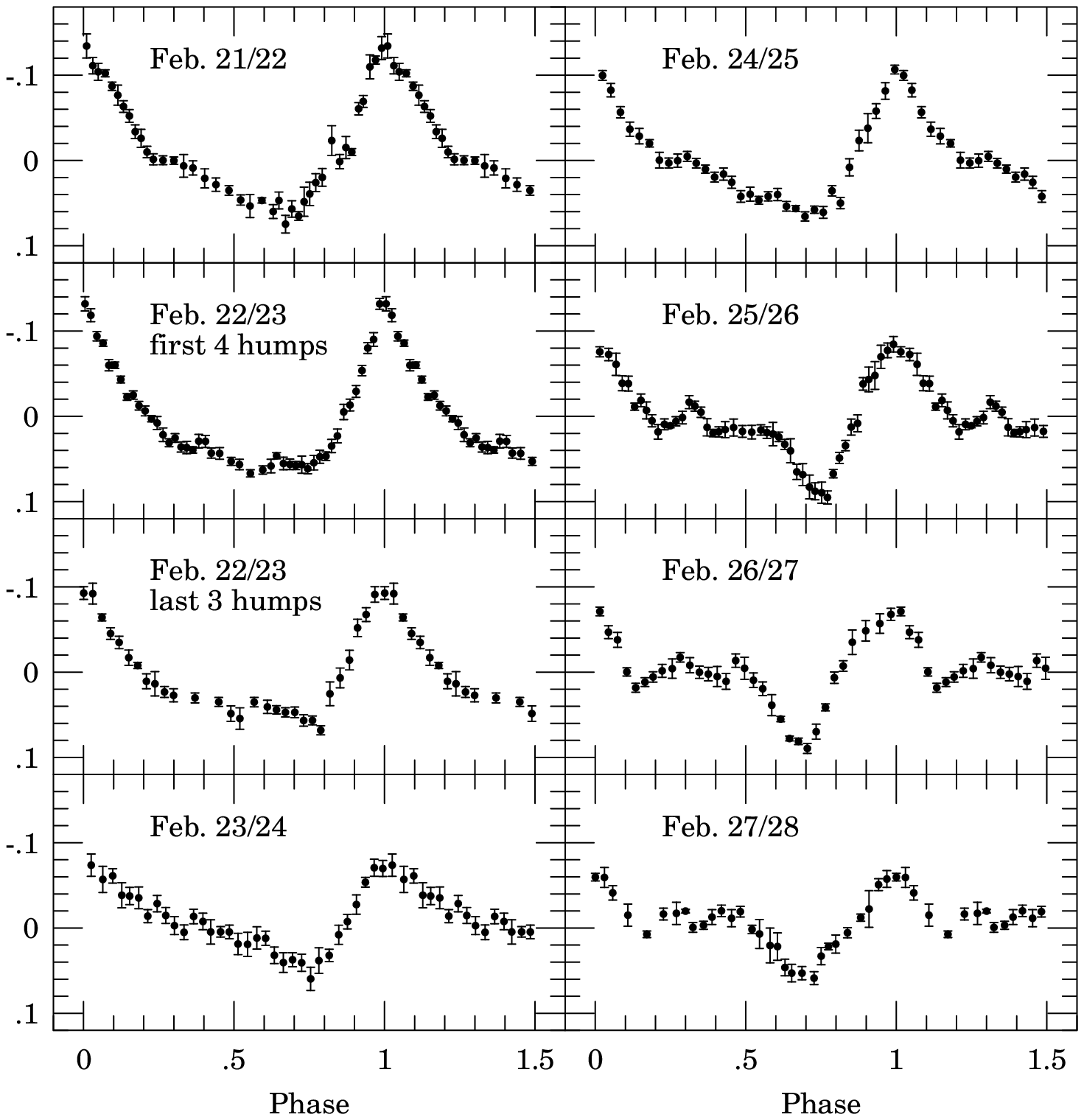}

\begin{figure}[h]
\caption{\sf Amplitude and profile changes of superhumps during 2003
superoutburst of KS UMa.}
\end{figure}
\medskip

The light curve of KS UMa in superoutburst was prewhitened with the main
period and its first harmonic. The power spectrum of the resulting light
curve shows no clear peaks except second, third and fourth harmonics of
the main frequency.

\clearpage
~
\vspace{7cm}

\includegraphics{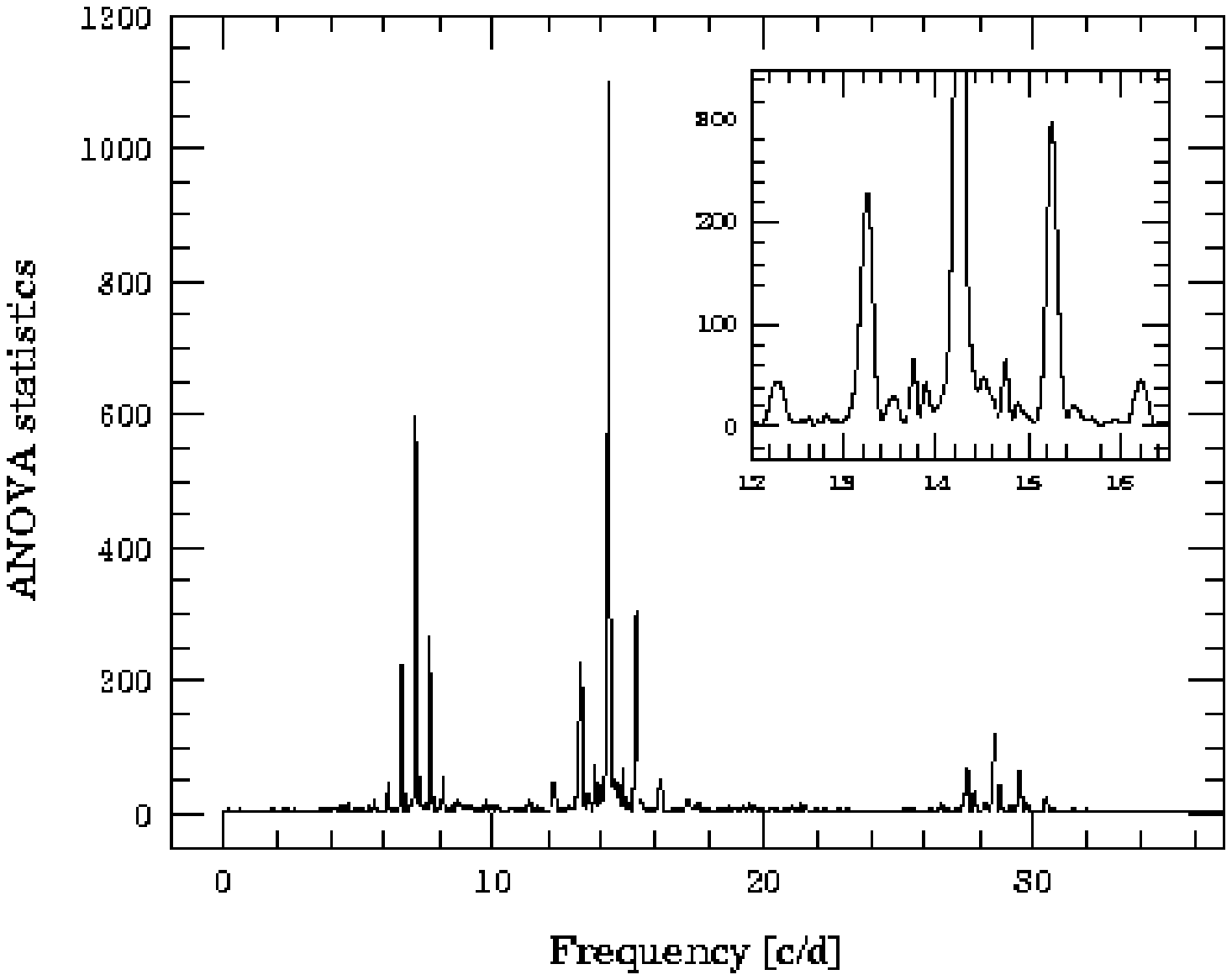}

\begin{figure}[h]
\caption{\sf {\sc anova} power spectrum of the light curve of KS UMa.
The inset shows the magnification of the power spectrum around main 
frequency.}
\end{figure}
\medskip

For nights from Feb. 21 to 27 we determined 27 times of maxima of
superhumps. They are shown in Table 2 together with their cycle numbers
$E$. The least squares linear fit to the data from Table 2 gives the
following ephemeris:

\begin{equation}
{\rm HJD}_{max} =  2452692.3279(12) + 0.070087(26) \cdot E
\end{equation}

\noindent indicating that the mean value of the superhump period
$P_{sh}$ is equal to 0.070087(26) days ($100.93\pm0.04$ min). This is in
good agreement with the value obtained from the power spectrum analysis.

Combination of both our determinations gives the mean value of superhump
period as equal to $P_{sh} = 0.070092(23)$ days.

\begin{table}[h]
\caption{\sc Times of maxima observed in the light curve
of KS UMa during its 2003 superoutburst}
\vspace{0.1cm}
\begin{center}
\begin{tabular}{|r|c|r||l|c|r|}
\hline
\hline
$E$ & ${\rm HJD}_{max}$ & $O - C$ & $E$ & ${\rm HJD}_{max}$ & $O - C$\\
    &             & [cycles] &     &             & [cycles]\\
\hline
 0 & 2452692.328 &  0.0014 & 43 & 2452695.342 &  0.0104\\
 3 & 2452692.538 & $-$0.0024 & 44 & 2452695.414 &  0.0305\\
 5 & 2452692.676 & $-$0.0335 & 45 & 2452695.486 &  0.0578\\
13 & 2452693.238 & $-$0.0153 & 56 & 2452696.258 &  0.0722\\
14 & 2452693.307 & $-$0.0308 & 57 & 2452696.326 &  0.0424\\
15 & 2452693.377 & $-$0.0321 & 58 & 2452696.392 & $-$0.0088\\
16 & 2452693.446 & $-$0.0477 & 59 & 2452696.464 &  0.0113\\
17 & 2452693.517 & $-$0.0347 & 60 & 2452696.536 &  0.0385\\
18 & 2452693.589 & $-$0.0074 & 73 & 2452697.448 &  0.0504\\
19 & 2452693.658 & $-$0.0230 & 74 & 2452697.514 & $-$0.0009\\
28 & 2452694.290 & $-$0.0060 & 85 & 2452698.280 & $-$0.0721\\
29 & 2452694.360 & $-$0.0073 & 86 & 2452698.350 & $-$0.0805\\
31 & 2452694.499 & $-$0.0241 & 87 & 2452698.420 & $-$0.0818\\
42 & 2452695.273 &  0.0188 &    &             &        \\
\hline
\hline
\end{tabular}
\end{center}
\end{table}
\bigskip

The $O - C$ departures from the ephemeris (1) are given also in Table 2
and shown in the lower panel of Fig. 8. The best fit to these data shown
as a solid line corresponds to the ephemeris:

\begin{equation}
{\rm HJD_{max}} = 2452692.3279 ~+~ 0.069804 E ~+~ 1.25 \times 10^{-5}~E^2
~-~ 1.15 \times 10^{-7}~E^3
\end{equation}
\hspace*{128pt} ${\pm}$0.0022 \hspace*{3pt} ${\pm}0.000233$ 
\hspace*{14pt} ${\pm}0.63$ \hspace*{56pt} ${\pm}0.47$\\
\bigskip

\bigskip

\subsection{Is KS UMa untypical?}

The upper panel of Fig. 8 shows the evolution of the peak-to-peak
amplitude of the superhumps during 2003 superoutburst of KS UMa. It
resembles in all details evolution of the superhump period shown in the
lower panel. Both the amplitude and the superhump period behavior looks
untypical for SU UMa stars. As we wrote earlier the amplitude of the
superhumps usually decreases monotonically during the superoutburst.

Till the mid of 1990ties all members of SU UMa group seemed to show only
negative superhump period derivatives (Warner 1985,  1995, Patterson et
al. 1993). It was interpreted as a result of disk shrinkage during the
superoutburst and thus lengthening its precession rate (Lubow 1992).
This picture become more complicated when the first stars with $\dot
P>0$ were discovered. Positive period derivatives were observed only in
stars with short superhump periods close to the minimum orbital period
for hydrogen rich secondary (e.g. SW UMa - Semeniuk et al. 1997a, WX Cet
- Kato et al. 2001a, HV Vir - Kato et. al 2001b) or for stars below this
boundary (e.g. V485 Cen - Olech 1997, 1RXS J232953.9+062814 - Uemura et
al. 2002).
 
The diversity of $\dot P$ behavior is well represented in the $\dot P/P$
versus $P_{sh}$ diagram shown in Fig. 9. This diagram is taken from Kato
et al. (2003a) with additional point for V1141 Aql (Olech 2003). It
shows also the position of the periodic gap and minimum period for
dwarf nova systems with hydrogen rich secondaries (Paczy\'nski 1981).
The outliers such as V485 Cen, 1RXS J232953.9+062814, KK Tel and TU
Men are also marked.

Recently, Nogami et al. (2003) reported observations of Var73 Dra - the
new SU UMa dwarf nova in the period gap. They found that the star having
mean superhump period of $P_{sh}=0.10623(16)$ days showed its change
rate of $\dot P/P = -1.7\pm0.2\times 10^{-3}$ which is one order of
magnitude larger than the largest values known. For clarity we do not
plot the position of Var73 Dra in our Fig. 9.

There are however few dwarf novae in which the superhump period
derivative is not constant and changes its sign. The first case of the
complex period behavior was observed during 1995 superoutburst of AL Com
(Howell et al. 1996) when during the first stage of superoutburst the
period was increasing with a rate of $\dot P/P=2.1\times 10^{-5}$ and
later decreased quite rapidly.

Complex behavior of period and amplitude of the superhumps was observed
recently in ER UMa (Kato et al. 2003b) and V1028 Cyg (Baba et al. 2000).
In the first stage of the outburst, superhump period of ER UMa was
increasing and around 5th day after superoutburst maximum ordinary
superhumps were switched into late superhumps which was connected with
change of the amplitude of the modulation and sign of the period
derivative. In the case of KS UMa the superhump period was also
increasing and changed its derivative between third and fourth night of
our observations (i.e. around fifth day after the maximum light as in
case of ER UMa). But as it is clearly visible from our Fig. 8 change of
the sign of period derivative in KS UMa was not connected with
transition to the late superhumps because we did not observe the
$\sim$0.5 phase shift in the superhumps maxima.

~

\vspace{15.8cm}

\includegraphics{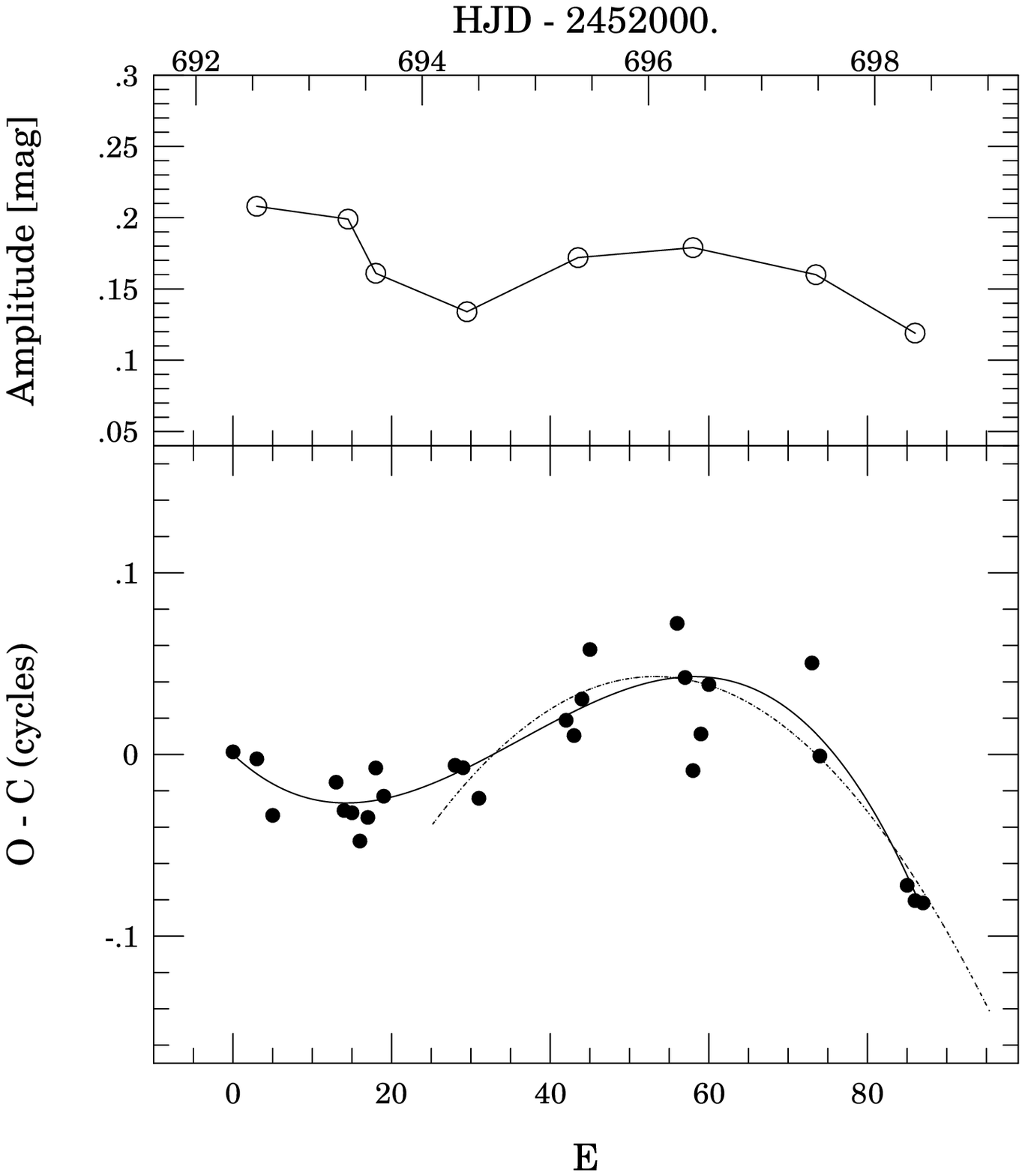}

\begin{figure}[h]
\caption{\sf {\it Upper panel}: Peak-to-peak amplitude changes of the
superhumps in KS UMa. ~{\it Lower panel}: O -- C diagram for 24 times
of superhump maxima of KS UMa. The solid line shows the relation computed
using ephemieris (2) and dotted line is a quadratic fit to the points
starting form cycle number 28.}
\end{figure}
\clearpage

~

\vspace{8.8cm}

\includegraphics{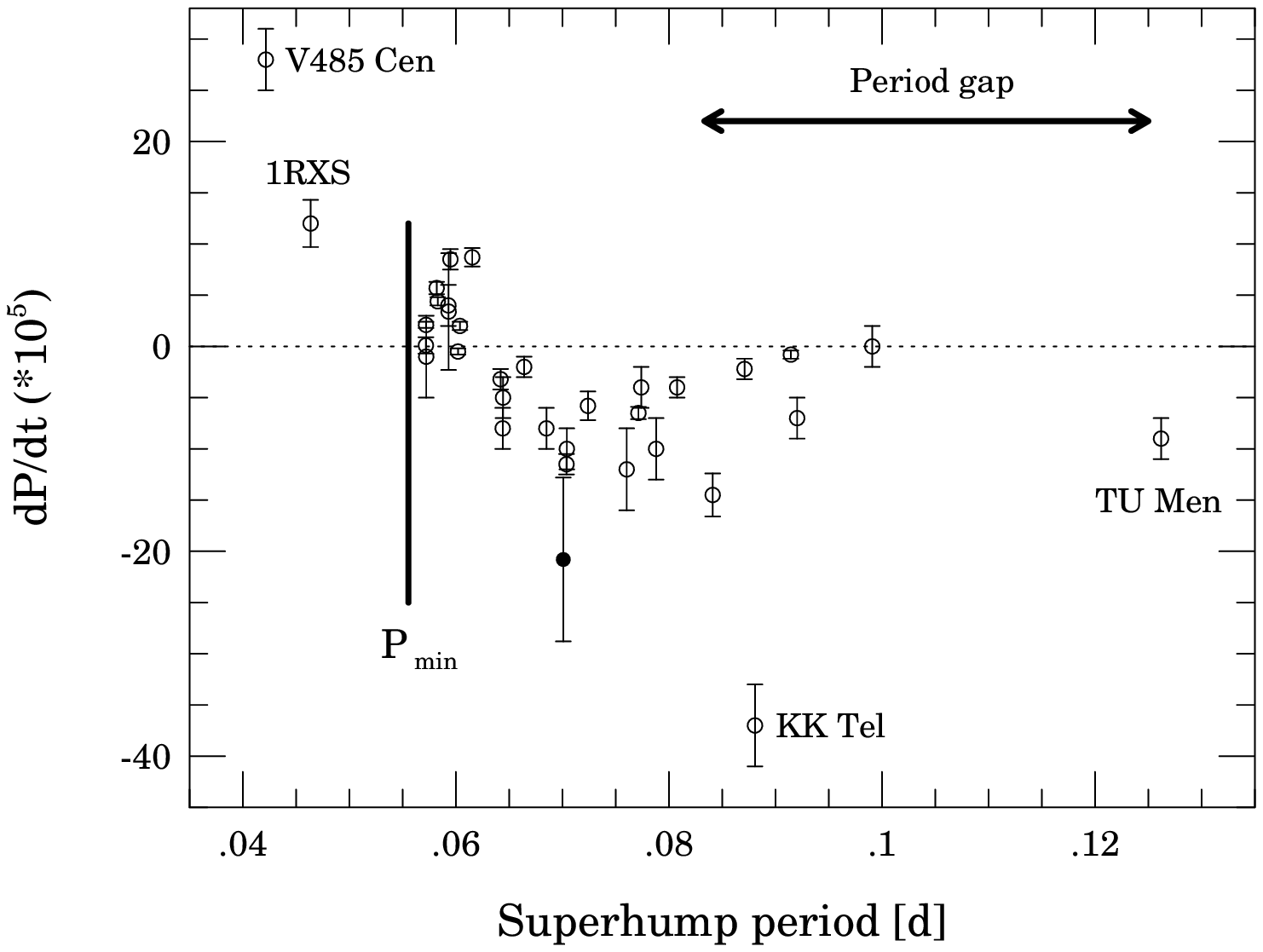}

\begin{figure}[h]
\caption{\sf $\dot P/P$ versus $P_{sh}$ for SU UMa-type dwarf novae.
The figure is taken from Kato et al. (2003a). Black dot shows the position
of KS UMa corresponding to the ephemeris shown by the dotted line in Fig. 8.
$P_{min}$ denotes the boundary for hydrogen rich secondary.}
\end{figure}
\medskip

Period and amplitude changes of KS UMa resemble closely those of V1028
Cyg during its 1995 superoutburst (Baba et al. 2000). In that case the
superhumps were fully developed on 1995 July 31. During the next six
days the amplitude decreased from 0.25 to 0.05 mag and the period was
increasing with the rate of $\dot P/P=8.7\times 10^{-5}$. Starting from
1995 Aug. 6 amplitude of the superhumps was larger again and period was
decreasing. The $O-C$ diagram shown by Baba et al. (2000) shows no signs
of 0.5 phase shift around Aug. 6 thus periodic light curve modulations
observed after this date are still ordinary not late superhumps.

According to Baba et al. (2000) V1028 Cyg may be a link between ordinary
SU UMa stars and WZ Sge subgroup of these variables. WZ Sge stars are
characterized by very long supercycles, large amplitudes of
superoutbursts and short orbital periods (close to 80 min boundary).
V1028 Cyg with orbital period around 87 min, supercycle over 400 days
and amplitude of the superoutburst around 6 mag is placed exactly
between ordinary SU UMa stars and WZ Sge variables. On the other hand it
this not the case of KS UMa whose supercycle is around one year, orbital
period around 98 min and amplitude of the outburst around 4 mag.

A possible explanation of the untypical behavior of KS UMa is assumption
that it is in fact quite typical. If we would start our observations
from night of Feb. 23/24, not two days earlier, we would conclude that
the period of the superhumps was decreasing with a rate of $\dot
P/P=-20\pm8 \times 10^{-5}$ (as is shown by dotted line in Fig. 8 and
filled circle in Fig. 9). 

Recent progress in development of cheap but quite sensitive CCD
detectors allowed astronomy amateurs to observe outbursts and detect
superhumps of many dwarf novae and collaborate with professional
astronomers. The excellent examples of such a fruitfull collaboration are
Center fo Backyard Astrophysics (CBA) run by Joseph Patterson from
Columbia University and also VSNET run by Taichi Kato and Daisaku
Nogami. Thus during last years we had usually very good coverage of
superoutburst of interesting objects and we have started to discover
such "peculiarities" as in case of KS UMa, V1028 Cyg, ER UMa or AL Com.
The question is when the number of such "peculiar" objects become so large
that we will begin to consider such a behavior as typical.

To check this hypothesis we reviewed the literature in search for
reported period variations in stars with $P_{sh}$ close to period of KS
UMa. The results of our search are summarized in Table 3 when we show
designation of the star, its mean superhump period, period derivative
in units of $10^{-5}$ and corresponding reference.

\begin{table}[h]
\caption{\sc Superhump period changes for stars with $P_{sh}$ close to
KS UMa}
\vspace{0.1cm}
\begin{center}
\begin{tabular}{llrl}
\hline
\hline
Object & $P_{sh}$ & $\dot P/P$ & Reference \\
\hline
V1028 Cyg & 0.06154 & $+8.7(0.9)$ & Baba et al. (2000)\\
V1159 Ori & 0.0642 & $-3.2(1)$ & Patterson et al. (1995)\\
CT Hya & 0.06643 & $-2(8)$ & Kato et al. (1999)\\
SX LMi & 0.0685 & $-8(2)$ & Nogami et al. (1997)\\
RZ Sge & 0.07039 & $-11.5(1)$ & Semeniuk et al. (1997b)\\
CY UMa & 0.0724 & $-5.8(1.4)$ & Harvey \& Patterson (1995)\\
\hline
\hline
\end{tabular}
\end{center}
\end{table}
\bigskip

The $O-C$ diagrams for stars from the papers listed in Table 3 are shown
in Fig. 10. The cycle numbers $E$ were renumerated to have $E\approx0$
corresponding approximately to the moment of birth of the superhump in
the light curve of the star.

What can we learn from Fig. 10? The most interesting thing is that this
figure forces us to revise previous statements on the superhump period
behavior. Warner (1985, 1995) and Patterson et al. (1993) argued that
the period derivative in SU UMa stars has a rather common negative value
of $\dot P/P \sim -5\times 10^{-5}$. From our Figs. 8 and 10 we can
clearly see that this is not true. Recently Kato et al. (2001b, 2003a)
indicated that most of long-period systems show a "textbook" decrease of
the superhump periods but short-period systems or infrequently
outbursting SU UMa type systems predominantly show an increase in the
superhump period. The transition between short and long period systems
is around period of 0.062 day thus V1028 Cyg with $P_{sh}=0.06154$ day
was short-period system and its $\dot P/P$ was positive while V1159 Ori
with $P_{sh}=0.0642$ day was included into a group of long-period
systems with negative $\dot P/P$ (as shown in Table 3).

Fortunately, observational coverage of the superoutbursts of V1028 Cyg
(Baba et al. 2000) and V1159 Ori (Patterson et al. 1995) was excellent
and comparison of the both $O-C$ shown in Fig. 10 indicates that in fact
at the beginning of the superoutburst the superhump period was
decreasing, in the middle phase of superoutburst was increasing and in
the third - the longest phase was again decreasing. Baba et al. (2002)
selected middle phase as representative for whole superoutburst of V1028
Cyg and obtained positive $\dot P/P$. On the other hand, Patterson et al.
(1995) simply fitted the parabola to all determined maxima of V1159 Ori
and therefore obtained negative value of $\dot P/P$.

The final conclusion of this section is that most probably all SU UMa
stars, both short and long period, show decreasing superhump period in
the beginning and in the end of the superoutburst but increasing period
in the middle phase. Our Fig. 10 proves it for medium and long period
systems. Recent observations of short period stars such as WZ Sge (see
Fig. 17 of Patterson et al. 2002), WX Cet (Fig. 7 of Kato et al. 2001a),
ER UMa (Fig. 2 of Kato et al. 2003b), EG Cnc (Fig. 5 of Patterson et al.
1998) and AL Com (Fig. 9 of Howell et al. 1996) suggest that it is also
true for this subgroup of SU UMa-type variables.

~

\vspace{19cm}

\includegraphics{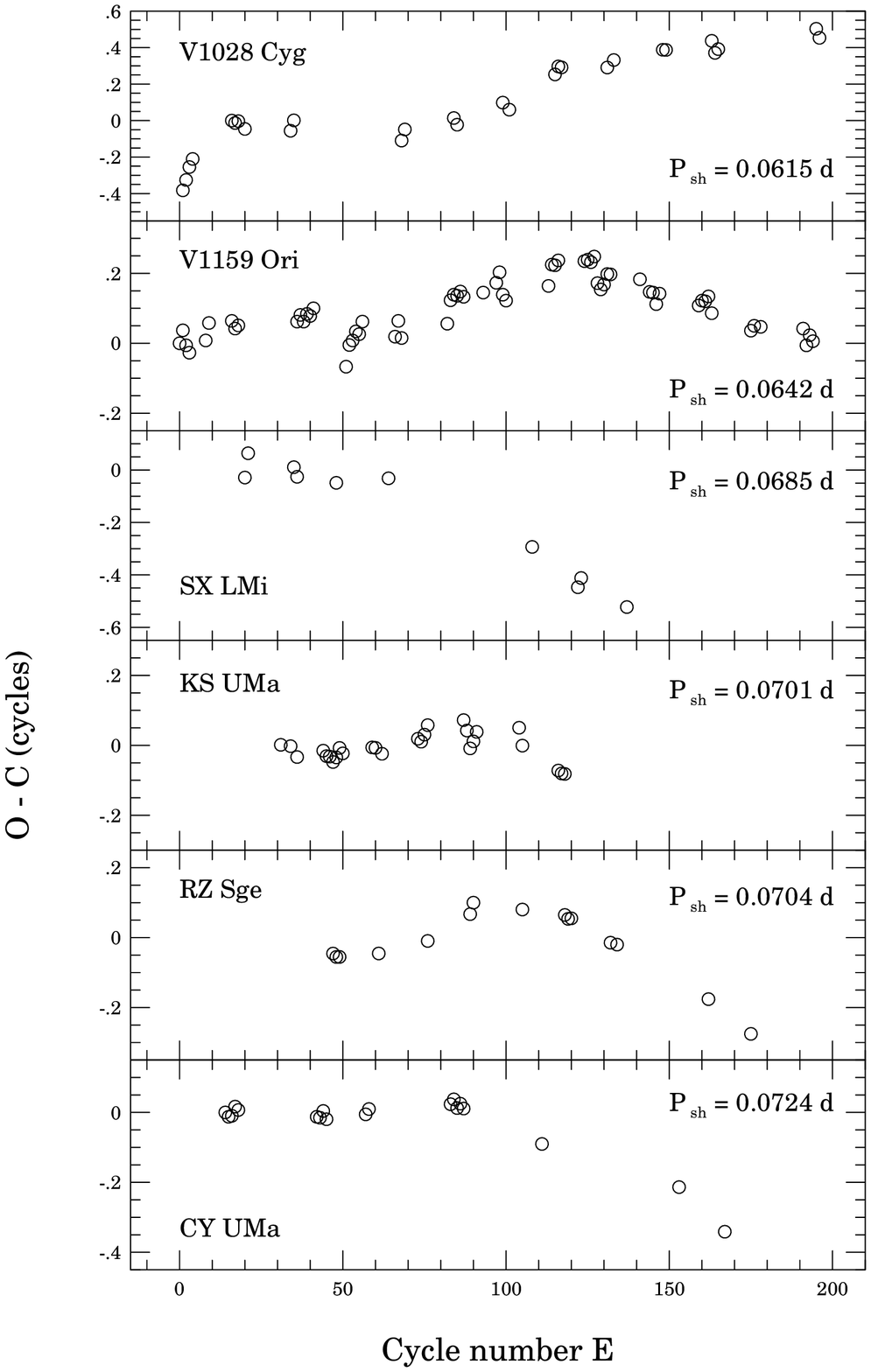}

\begin{figure}[h]
\caption{\sf O -- C diagrams for SU UMa-type dwarf novae with superhump 
period close to 0.07 days.}
\end{figure}
\clearpage

In the case of KS UMa during the first four days of our observations
(days 4 -- 7 of superoutburst) the period of superhumps was increasing
with a rate of $\dot P/P = (21\pm12)\times 10^{-5}$ and later (days 7 --
12 of superoutburst) was decreasing with a rate of $\dot P/P =
-(21\pm8)\times 10^{-5}$.

\section{Late superhumps}

The 2003 superoutburst of KS UMa lasted until March 06/07 but modulations
with period close to $P_{sh}$ were observed even till March 22/23 (see
Fig. 4). The shape and amplitude of these modulations was changing very
quickly, sometimes from cycle to cycle, thus for searching for
periodicities we decided to use ordinary Fourier transform. The power
spectrum for period from March 05/06 to 21/22 is shown in Fig. 11. The
highest peak is found at the frequency of $14.38\pm0.020$ which
corresponds to the period of $0.0695\pm0.0001$ days.

~

\vspace{11cm}

\includegraphics{fig11.ps}

\begin{figure}[h]
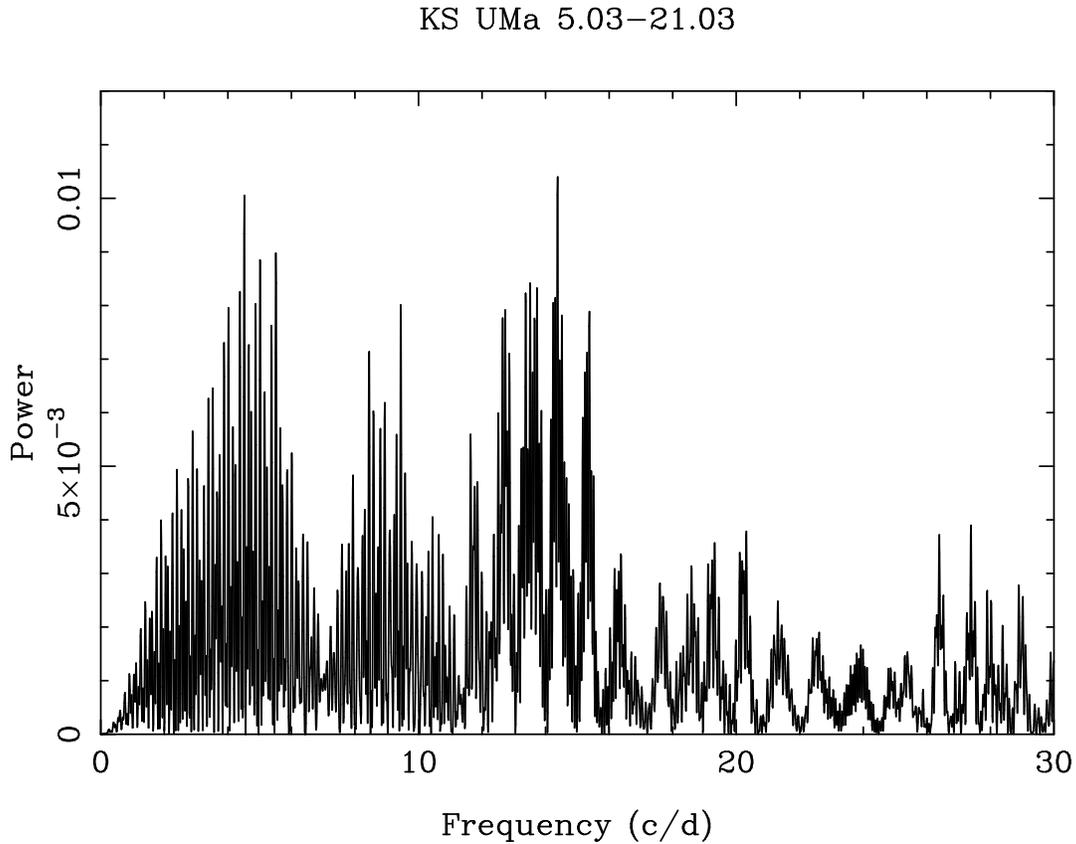

\caption{\sf Power spectrum of the light curves of KS UMa from
March 05/06 to 21/22}
\end{figure}
\bigskip

In the case of these late superhumps the minima were more clear and
sharp than maxima thus were much better for $O-C$ analysis. Finally, in
period Mar. 06 -- 22 we determined 11 times of minima, which are listed
in Table 4 together with their cycle counts $E$ and $O-C$ values
computed according to the following ephemeris:

\begin{equation}
{\rm HJD}_{min} = 2452706.4341(20) ~+~ 0.06925(2)\cdot E
\end{equation}

\begin{table}[h]
\caption{\sc Times of minima observed in the light curve
of KS UMa during late superhump stage}
\vspace{0.1cm}
\begin{center}
\begin{tabular}{|r|c|r||l|c|r|}
\hline
\hline
$E$ & ${\rm HJD}_{min}$ & $O - C$ & $E$ & ${\rm HJD}_{min}$ & $O - C$\\
    &             & [cycles] &     &             & [cycles]\\
\hline
 $-17$ & 2452705.258 &  0.0166 & 201 & 2452720.353 &  $-0.0051$ \\
 $-2$ & 2452706.295 &  $-0.0086$ & 202 & 2452720.415 &  $-0.1097$ \\
 $-1$ & 2452706.364 & $-0.0124$ & 203 & 2452720.474 & $-0.2577$\\
    0 & 2452706.435 & 0.0130 & 204 & 2452720.564 & 0.0419\\
    1 & 2452706.504 & 0.0094 & 217 & 2452721.480 & 0.2693\\
  200 & 2452720.277 & $-0.1025$ & & & \\
\hline
\hline
\end{tabular}
\end{center}
\end{table}
\bigskip

$O-C$ deviations are sometimes large but show no clear trend. We thus
conclude that period of the late superhumps was roughly constant and
its value was (combination of $O-C$ and power spectrum estimates):
$P_{late} = 0.06926(2)$ days.

\vspace{11cm}

\includegraphics{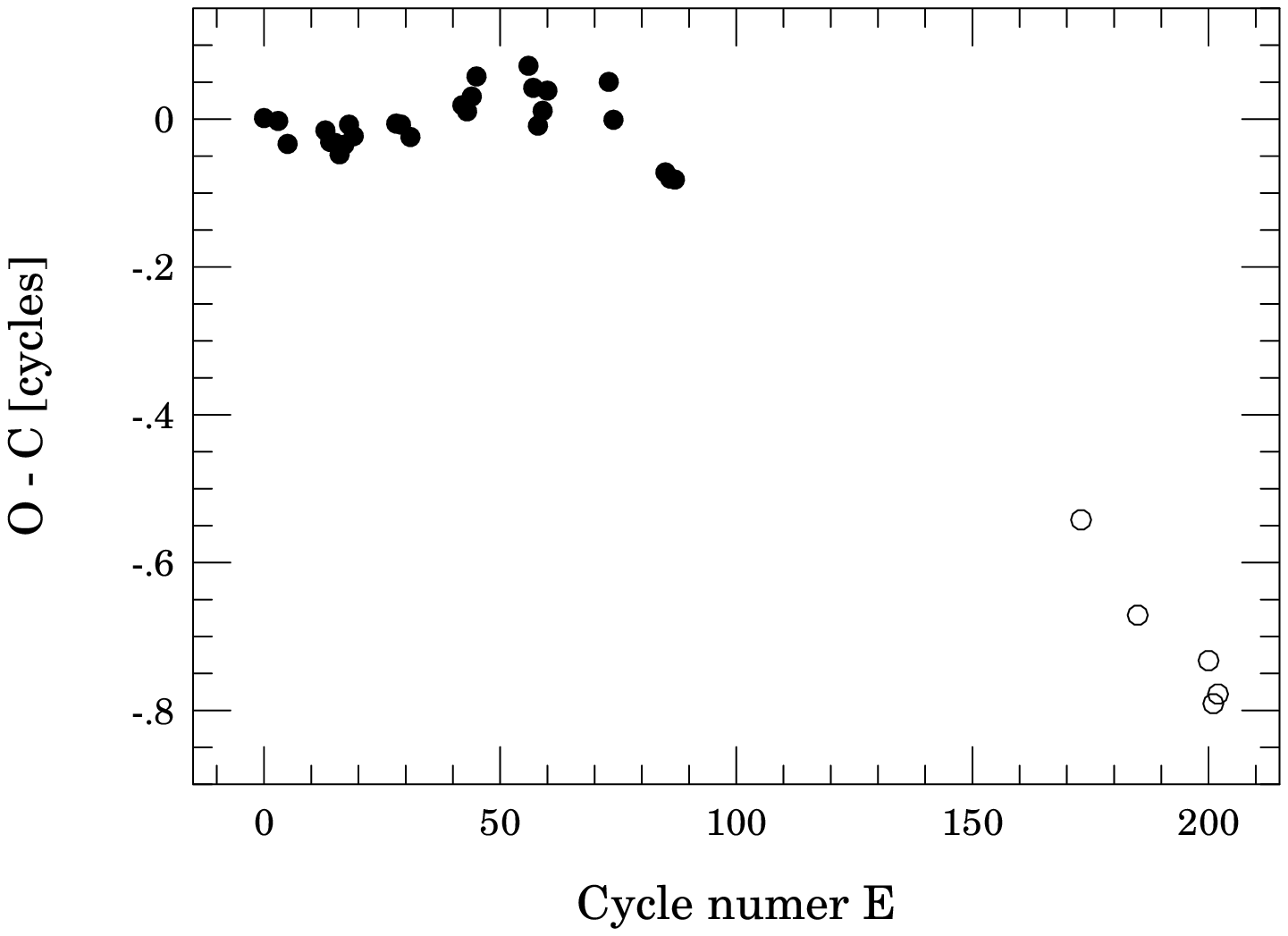}

\begin{figure}[h]
\caption{\sf O -- C diagram for common (filled circles) and late 
(open circles) superhump maxima of KS UMa. O -- C departures are
computed using ephemeris (1).}
\end{figure}

Despite of the five day break in observations between Feb. 28 and Mar. 4
we decided to determine the values of the maxima in the light curve of
KS UMa in early March and connect them with the maxima observed in
February. The March maxima are listed in Table 5 together with cycle
numbers $E$ and $O-C$ departures computed according to the ephemeris
(1). The $O-C$ values are shifted in phase by 0.5.

\begin{table}[h]
\caption{\sc Times of maxima observed in the light curve
of KS UMa during late superhump stage}
\vspace{0.1cm}
\begin{center}
\begin{tabular}{|r|c|r||l|c|r|}
\hline
\hline
$E$ & ${\rm HJD}_{max}$ & $O - C$ & $E$ & ${\rm HJD}_{max}$ & $O - C$\\
    &             & [cycles] &     &             & [cycles]\\
\hline
    173 & 2452704.450 & $-0.5421$ & 201 & 2452716.395 & $-0.7909$\\
    185 & 2452705.282 & $-0.6711$ & 202 & 2452716.466 & $-0.7779$\\
    200 & 2452706.329 & $-0.7326$ &     &             &          \\
\hline
\hline
\end{tabular}
\end{center}
\end{table}
\bigskip

The $O-C$ departures for common and late superhumps are shown in Fig.
12. Provided our cycle count is correct, we can conclude that in the
second stage of the superoutburst and during late superhump phase the
superhump period decreased with a rate of $\dot
P/P=-(6.0\pm1.1)\times10^{-5}$ i.e. value quite typical for medium and
long period SU UMa stars.

\section{Quiescence}

As displayed in Fig. 5 KS UMa in quiescence shows quasi-periodic
modulations with amplitude reaching even 0.5 mag. The most
characteristic feature observed in this stage was sinusoidal wave with
period around 0.1 days clearly visible during late March nights. 

The Fourier power spectrum for nights Mar. 24/25 -- Apr. 01/02 is shown
in Fig. 13. Before calculation the light curves were prewhitened using
second order polynomial.

The periodogram yields no specific frequency of these modulations. The
highest peak found in Fig. 13 corresponds to the frequency
$10.20\pm0.02$ c/d i.e. to the period of $0.0980\pm0.0002$ days and
exceeds only marginally compeating features. 

\section{Summary}

We reported extensive photometry of the dwarf nova KS UMa in its 2003
superoutburst and quiescence. The amplitude of the superoutburst was 3.9
mag. The maximal brightness of the star was 12.3 mag and the mean
magnitude in quiescence was 16.2.

The 2003 superoutburst of KS UMa lasted from Feb. 18/19 to Mar. 06/07 i.e.
16 days. During this interval the star showed clear superhumps with a
mean period of $P_{sh} = 0.070092(23)$ days. 

On Feb. 21/22 the amplitude of the superhumps was 0.21 mag and decreased
to 0.13 mag on Feb. 23/24. Surprisingly, later amplitude increased 
reaching its local maximum with 0.18 mag on Feb. 25/26.

In the middle stage of superoutburst the period of superhumps was
increasing with a rate of $\dot P/P = (21\pm12)\times 10^{-5}$ and later
was decreasing with a rate of $\dot P/P = -(21\pm8)\times 10^{-5}$.
Comparing KS UMa to other SU UMa stars we concluded that this group of
dwarf novae shows decreasing superhump periods at the beginning and the
end of superoutburst but increasing period in the middle phase. This is
contrary to the original suggestion of Warner (1985, 1995) and Patterson
et al. (1993) that superhump periods usually decrease with a rate around
$-5\times10^{-5}$ and also in contrast with a recent investigation of
Kato et al. (2001a) who concluded that short period systems shows
increasing periods but long period SU UMa stars are characterized by
decreasing periods.

\vspace{11.5cm}

\includegraphics{fig13.ps}

\begin{figure}[h]
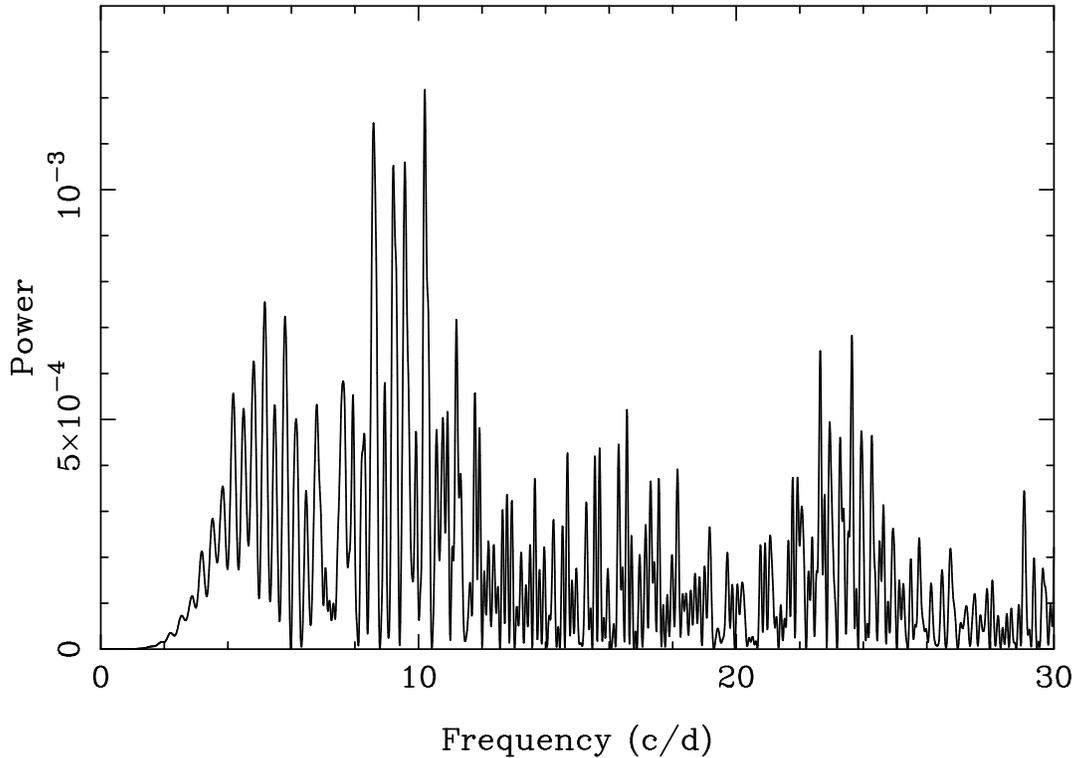

\caption{\sf Power spectrum of the light curves of KS UMa from
Mar. 24/25 to Apr. 01/02}
\end{figure}
\bigskip

At the end of the superoutburst and during first dozen days of quiescence
the star showed late superhumps with a mean period of $P_{late} =
0.06926(2)$ days. This phenomenon was observed even 30 days after
beginning of the superoutburst.

In quiescence the star shows quasi-periodic modulations with amplitude
reaching 0.5 mag. The most common structure observed during this stage
was sinusoidal wave characterized by period of 0.098 days.

\bigskip \noindent {\bf Acknowledgments.} ~We acknowledge generous
allocation of  the Warsaw Observatory 0.6-m telescope time. This work
was partially supported by KBN grant number 2~P03D~002~20 and used
the online service of the VSNET. We would like to than Prof. J\'ozef Smak
for reading and commenting on the manuscript.

\end{document}